\documentclass[aps,prl,
superscriptaddress,
twocolumn,
floatfix,
amsmath,amssymb,amsfonts,
letterpaper,
]{revtex4-1}

\usepackage{graphicx}
\usepackage{hyperref}
\usepackage{color}
\usepackage{array}

\begin{document}

\title{A generic, hierarchical framework for massively parallel Wang--Landau sampling}

\author{Thomas Vogel}
\email{thomasvogel@physast.uga.edu}
\affiliation{Center for Simulational Physics, The University of
  Georgia, Athens, GA 30602, USA}
\author{Ying Wai Li}
\affiliation{Center for Simulational Physics, The University of Georgia, Athens, GA 30602, USA}
\affiliation{National Center for Computational Sciences, Oak Ridge
  National Laboratory, Oak Ridge, TN 37831, USA}
\author{Thomas W\"ust}
\affiliation{Swiss Federal Research Institute
  WSL, Z\"{u}rcherstrasse 111, CH-8903 Birmensdorf, Switzerland}
\author{David P. Landau}
\affiliation{Center for Simulational Physics, The University of
  Georgia, Athens, GA 30602, USA}

\begin{abstract}
\noindent
We introduce a parallel Wang--Landau method based on the
replica-exchange framework for Monte Carlo simulations. To demonstrate
its advantages and general applicability for simulations of complex
systems, we apply it to different spin models including spin glasses,
the Ising model and the Potts model, lattice protein adsorption, and
the self-assembly process in amphiphilic solutions. Without loss of
accuracy, the method gives significant speed-up and potentially scales
up to petaflop machines.
\end{abstract}


\maketitle

In Wang--Landau (WL) sampling, the \emph{a priori} unknown density of
states $g(E)$ of a system is iteratively determined by performing a
random walk in energy space ($E$) seeking to sample configurations
with probability $1/g(E)$ (``flat
histogram'')~\cite{wl_prl,wl_pre,shanho04ajp}. This procedure has
proven very powerful in studying problems with complex free energy
landscapes by overcoming the prohibitively long time scales typically
encountered near phase transitions or at low temperatures. It also
allows us to calculate thermodynamic quantities, including the free
energy, at any temperature from a single simulation. Moreover,
Wang--Landau sampling is a generic Monte Carlo procedure with only a
minimal set of adjustable parameters and, thus, has been applied
successfully to such diverse problems as spin glasses, polymers,
protein folding, lattice gauge theory, etc.,
see~\cite{rathore02jcp,alder04jsm,taylor09jcp,langfeld12prl} for
examples. Various improvements have been proposed to the method,
either by optimizing the ``modification factor--flatness criterion''
scheme~\cite{zhou05pre,zhou06prl,belardinelli07pre} or by means of
efficient Monte Carlo trial
moves~\cite{yamaguchi02pre,wu05pre,wuest09prl} (to name a few).
Ultimately, however, parallelization is the only means to
systematically sustain the performance for ever larger problems.
Surprisingly, to date, only two directions have been taken in this~\hbox{regard}:

Parallelization scheme (i): As already
suggested~\cite{wl_pre,shanho04ajp}, it is possible to subdivide the
total energy range into smaller sub-windows each sampled by an
independent WL instance (random walker). The total simulation time is
obviously limited by the convergence of the slowest walker and can be
tuned by unequal distribution of energy space. However, an optimal
load balancing is impossible due to the \emph{a priori} unknown
irregularities in the complex free energy landscape. Moreover, energy
intervals cannot be reduced arbitrarily due to systematic errors
introduced from ``locked-out'' configurational space.

Parallelization scheme (ii): Here, multiple random walkers work
simultaneously on the \emph{same} density of states (and histogram).
Distributed memory (MPI~\cite{khan05jcp}), shared memory
(OpenMP~\cite{zhan}), and GPU~\cite{yin12cpc} variants of this idea
have been proposed; shared memory implementations have the advantage
of not requiring periodic synchronization among the walkers and even
allowing for ``data race'' when updating $g(E)$ without noticeable
loss in accuracy~\cite{zhan}.  Although this second approach seemingly
avoids the problems of scheme~(i), a~recent, massively parallel
implementation~\cite{yin12cpc} has revealed that correlations among
the walkers can systematically underestimate the DOS in hardly
accessible energy regions. A~remedy to the problem has been proposed
in terms of a (heuristic) bias to the modification factor; but,
overall, such inter-dependencies render this parallelization scheme
highly problematic. Moreover, it is important to note that the
effective round-trip times of the individual walkers are not improved
by this concerted update.

In this Letter, we propose a \emph{generic} parallel Wang-Landau
scheme which combines the advantageous dynamics of Wang-Landau
sampling with the idea of replica-exchange Monte
Carlo~\cite{partemp1,partemp3}.  Similar to scheme~(i), we start off
by splitting up the total energy range into smaller sub-windows but
with large overlap between adjacent windows.  Each energy sub-window
is sampled by multiple, \textit{independent} WL walkers.  The key to
our approach is that configurational or replica exchanges are allowed
among WL instances of overlapping energy windows during the course of
the simulation, such that each replica can travel through the entire
energy space. The replica exchange move does not bias the overall WL
procedure and, thus, guarantees the flexibility to be applied to any
valid WL update/convergence rule (e.g. the $1/t$
algorithm~\cite{belardinelli07pre}).  Furthermore, our hierarchical
parallelization approach does not impose any principal limitation to
the number of WL instances used [contrary to scheme (i), see above].
Therefore, it is conceivable to design setups which scale up to
thousands of~CPUs.


The standard WL algorithm~\cite{wl_prl,wl_pre} estimates the density
of states, $g(E)$, in an energy range $\left[ E_{\textrm{min}},
  E_{\textrm{max}}\right]$ using a single random walker. During the
simulation, trial moves are accepted with a probability $P =
\textrm{min} \left[1, g(E_\textrm{old})/g(E_\textrm{new})\right]$,
where $E_\textrm{old}$ ($E_\textrm{new}$) is the energy of the
original (proposed) configuration. The estimation of $g(E)$ is
continuously adjusted and improved by a modification factor $f$ (as
$g(E)\to f\times g(E)$) which gets progressively closer to unity as
the simulation proceeds, while a histogram $H(E)$ keeps track of the
number of visits to each energy $E$ during an iteration.  When $H(E)$
is sufficiently ``flat'', the next iteration begins with $H(E)$ reset
to zero and $f$ reduced by some predefined rule (e.g. $f \rightarrow
\sqrt f$). The simulation terminates when $f$ reaches a small enough
$f_\textrm{final}$ at which point the accuracy of $g(E)$ is
proportional to $\sqrt f_\textrm{final}$ for flat enough
$H(E)$~\cite{zhou05pre}.

\begin{figure}
  \includegraphics[width=\columnwidth,clip]{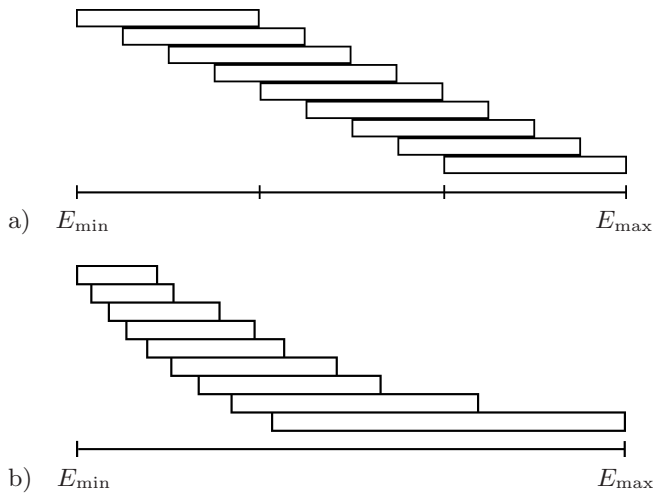}
  \caption{\label{fig:method_1}%
    a) Partition of the global energy range into nine equal-size
    intervals with overlap $o = 75\%$. b) Run-time balanced partition
    with overlap to the higher energy interval $o\geq75\%$. Multiple
    WL walkers can be employed in each interval.}
\end{figure}

In our parallel WL scheme, the global energy range is
first split into $h$ smaller intervals (sub-windows), each of which
contains $m$ random walkers. Consecutive intervals must overlap each
other to allow for configurational exchange, see
Fig.~\ref{fig:method_1} for examples.  The overlap $o$ should be
neither too large nor too small so as to strike a balance between fast
convergence of $g(E)$ and a reasonable exchange acceptance rate. In
fact, we find that a large overlap of $o\approx75\%$ is advantageous,
but that number is flexible to a certain extend and one can also
obtain excellent results with other choices~\cite{tv_follow_2013}.
Within an energy sub-window, each random walker performs standard WL
sampling.  After a certain number of Monte Carlo steps, a replica
exchange is proposed between two random walkers, $i$ and $j$, where
walker $i$ chooses swap partner $j$ from a neighboring window at
random. Let $X$ and $Y$ be the configurations that the random walkers
$i$ and $j$ are carrying before the exchange; $E(X)$ and $E(Y)$ be
their energies, respectively. From the detailed balance condition the
acceptance probability $P_{\text{acc}}$ for the exchange of
configurations $X$ and $Y$ between walkers $i$ and $j$ is:
\begin{equation}
  P_{\text{acc}}=\min\left[1,\frac{g_i(E(X))}{g_i(E(Y))}\frac{g_j(E(Y))}{g_j(E(X))}\right]\,,
\end{equation}
where $g_i(E(X))$ is the instantaneous estimator for the density of states
of walker $i$ at energy $E(X)$, cf.~\cite{nogawa11pre}.

In contrast to parallelization scheme (ii) above, in our formalism,
every walker is furnished with its own $g(E)$ and $H(E)$ which are
updated independently. Also, every walker has to fulfill the WL
flatness criterion independently at each iteration, ensuring that
systematic errors as found in~\cite{yin12cpc} cannot occur.  When all
random walkers within an energy sub-window have individually attained
flat histograms, their estimators for $g(E)$ are averaged out and
redistributed among themselves before simultaneously proceeding to the
next iteration.  This practice reduces the error during the simulation
with ${\sqrt{m}}$~\cite{tv_follow_2013}, i.e. as for uncorrelated WL
simulations. Furthermore, increasing $m$ can improve the convergence
of the WL procedure by reducing the risk of statistical outliers in
$g(E)$ resulting in slowing down subsequent iterations.
(Alternatively, it allows us, in principle, to use a weaker flatness
criterion~\cite{tv_follow_2013}, which is in the spirit of a
concurrently proposed idea of merging histograms in multicanonical
simulations~\cite{zierenberg13cpc}.)  The simulation is terminated
when all the energy intervals have attained $f_\textrm{final}$. At the
end of the simulation, $h\times m$ pieces of $g(E)$ fragments with
overlapping energy intervals are used to calculate a single $g(E)$ in
the complete energy range. During that procedure, the joining point
for any two overlapping density of states pieces is chosen where the
inverse microcanonical temperatures
$\beta=\textrm{d}\log[g(E)]/\textrm{d}E$ best coincide, and
statistical errors are determined by resampling
techniques~\cite{newman_book,tv_follow_2013}.

\begin{figure}
  \includegraphics[width=\columnwidth,clip]{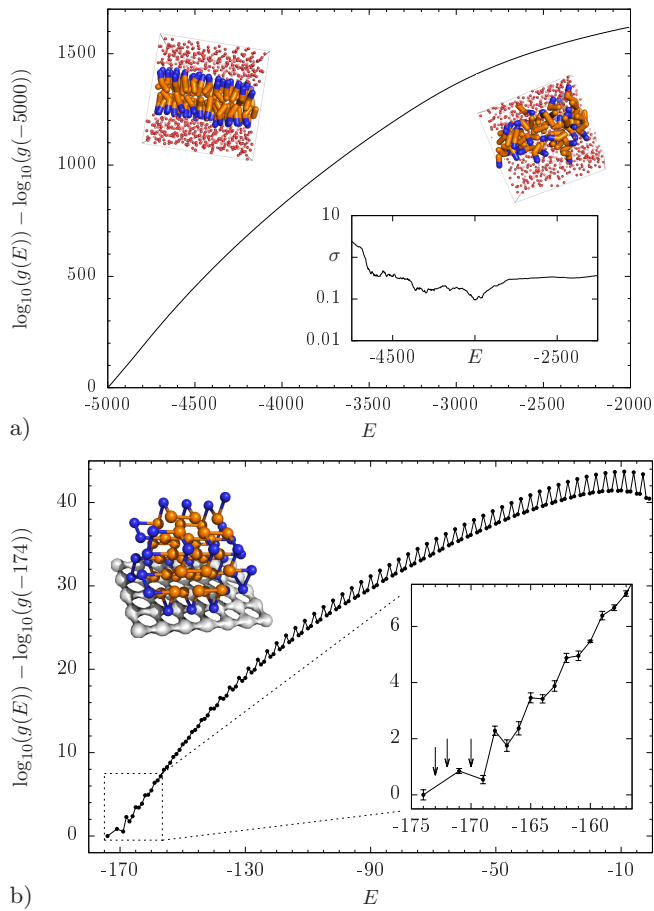}
  \caption{\label{fig:two_dos}%
    Logarithm of density of states (DOS) obtained by our parallel WL scheme with
    the setup shown in Fig.~\ref{fig:method_1}\,a. (Top) Amphiphilic
    system containing 75 lipid molecules and a total of 1000
    particles. Error bars ($\sigma$), obtained from multiple independent
    simulations, are smaller than the line thickness and shown in the inset.  The pictures show a conformation
    where lipid molecules assemble and form a single cluster
    ($E\approx-2100$) and a low-energy bilayer configuration
    ($E\approx-4800$).\break{}
    (Bottom) DOS of the lattice HP 67mer, where only H-monomers are
    attracted by the substrate. The H-H interaction is 3 times
    stronger than the surface attraction leading to the unusual
    sawtooth like shape. The inset shows the error bars on the
    enlarged low-energy data. Note the two energy gaps, i.e. no
    conformations exist with $E=-173$, $-172$, and $-170$
    (see arrows). The picture shows an adsorbed HP protein
    with energy $E=-174$.}
\end{figure}
In order to assess its general applicability, feasibility and
performance, we applied this novel parallel WL scheme to multiple
models in statistical mechanics. The first two are the well studied
Ising model and 10-state Potts model in 2 dimensions, showing
second-order and first-order transitions, respectively. We applied the
parallel scheme to the 2D Ising model up to system sizes of $256^2$
using up to the order of 2000 cores.  The deviations from exact
results were always of the same order as the statistical errors, which
are $<0.01\%$ in the peak region of the density of states. For the
10-state Potts model, we extrapolated the critical temperature in the
thermodynamic limit from results of system sizes up to $300^2$. Our
estimate of $T_\mathrm{c}^\infty=0.701234\pm0.000006$ is in excellent
agreement with the exact value of $0.701232$~\cite{baxter73jpc}.
While it still takes a few days for a single-walker WL run to converge
for a $100^2$ Potts system, we obtained all results, which will be
shown in detail elsewhere~\cite{tv_follow_2013}, within hours. Besides
this remarkable accuracy and absolute gain in timing, we will show
below that our algorithm has almost perfect weak scaling behavior for
these lattice models since their system sizes are scalable in a
straightforward way. For a final test, we applied the method to a
$12\times12\times12$ spin-glass system and reproduced results
published earlier~\cite{berg93epl,wl_pre} with speed-ups of the same
order as reported below. In particular, low energy states are found
much faster, while our estimate for $e_0=-1.787\pm0.005$ agrees
perfectly with the earlier data~\cite{wl_pre}. To demonstrate the
potential to obtain new physics results and strong scaling properties,
we also apply the method to two very distinct and particularly
challenging molecular problems: a coarse-grained continuum model for
the self assembly of amphiphilic molecules (lipids) in explicit
solution and a lattice model for the surface adsorption of proteins.
In the first model, amphiphilic molecules, each of which composed of a
polar (P) head and two hydrophobic (H) tail monomers (P--H--H), are
surrounded by solvent particles (W).  Interactions between H and W
molecules, as well as those between H and P molecules, are purely
repulsive. All other interactions between non-bonded particles are of
Lennard-Jones type; bonded molecules are connected by a FENE
potential, cf.~\cite{getz,fujiwara} for similar models. The second
model uses the hydrophobic-polar (HP) model~\cite{Dill1985} for
protein surface adsorption. Here a protein is represented by a
self-avoiding walk consisting of H and P monomers placed on a cubic
lattice with an attractive substrate. Recent studies on this model and
details can be found in~\cite{Li2011,Li2012}.

Both models bring about qualitatively different technical challenges,
such as high energy and$/$or configurational barriers, and simulations
of particular setups are \emph{impossible} for all practical purposes
using the traditional, single walker WL method due to unreasonable
resource demands. For a demonstration, we choose two such systems. The
first consists of 75 lipid molecules and 775 solution particles using
the first model with a continuous energy domain.  The density of
states $g(E)$ on an energy range covering the lipid bilayer formation
spans more than 1600 orders of magnitude (cf.
Fig.~\ref{fig:two_dos}\,a), which makes low temperature statistics
extremely difficult to obtain.  The second system is an HP lattice
protein consisting of 67 monomers~\cite{yue95pnas} interacting with a
weakly attractive surface and with discrete energy levels, which gives
rise to an unusually rugged density of states, see
Fig.~\ref{fig:two_dos}\,b.  \hbox{Obtaining} convergence for the
entire energy range is an arduous task using a single walker.

\begin{figure}
  \includegraphics[width=\columnwidth,clip]{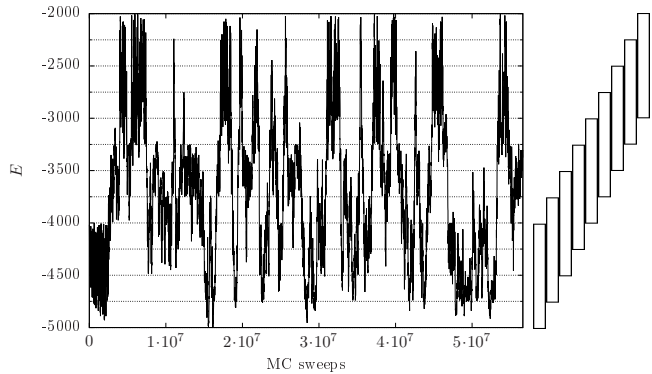}
  \caption{\label{fig:wanderlogs}%
    Path of a single replica through energy space. Replica exchange
    between walkers is proposed every $10^4$ sweeps (data also shown
    with that resolution), with acceptance rates between 30 and
    55\,\%. Grid lines correspond to the borders of the individual
    energy windows, cf. sketch of parallel setup at right and
    Fig.~\ref{fig:method_1}\,a.}
\end{figure}

Our parallel WL framework allows us to successfully simulate both,
previously inaccessible, systems. The reasons are two-fold: first,
each walker is now responsible for sampling a smaller configurational
phase space, which contributes mainly to the faster convergence.
Second, the replica exchange process revitalizes walkers from trapped
states and avoids an erroneous bias in $g(E)$ due to potential
ergodicity breaking since replicas can access the \textit{entire}
conformational space by walking through all energy windows. A typical
time-series of a replica performing round trips in the full energy
range of the lipid system (cf.  Fig.~\ref{fig:two_dos}\,a) is shown in
Fig.~\ref{fig:wanderlogs}. With these features combined, we obtain the
entire $g(E)$ with a noticeable speed-up and high accuracy,
see~\cite{tv_follow_2013} for more details.

\begin{figure}
  \includegraphics{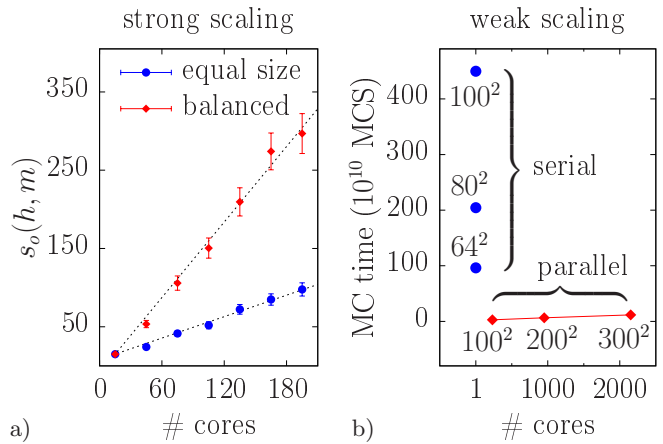}
  \caption{\label{fig:scaling} a) Speed-up $s_o(h,m)$ for different
    numbers of cores for equal-size energy windows and
    overlap $o = 75\%$ (blue circles, cf.  Fig.~\ref{fig:method_1}\,a)
    and using a run-time balanced energy splitting (red diamonds, cf.
    Fig.~\ref{fig:method_1}\,b). Here, the calculation of the speed-up
    is based on the MC steps (MCS) needed to complete the
    first WL iteration. b) MC time to terminate serial WL
      runs for different system sizes of the 2D Potts model (blue
      circles) vs. run time for parallel runs if the number of cores
      increases according to the increase in system size (red
      diamonds). The run time practically stays constant, proving the
      weak scaling property of our method.}
\end{figure}

To quantify the efficiency of the parallel WL scheme, we define the
speed-up, $s_o(h,m)$, as the number of Monte Carlo steps taken by the
slowest parallel WL walker ($N_o^{\mathrm{parallel}}(h,m)$), as compared
to that taken by $m$ single walkers ($N^{\mathrm{single}}$):
\begin{equation}
\label{eq:speed}
s_o(h,m) =\frac{N^{\mathrm{single}}}{N_o^{\mathrm{parallel}}(h,m)}\,.
\end{equation}
For $h \lesssim 20$ we have achieved strong scaling: the speed-up
scales \textit{linearly} with the increase in $h$ as shown in
Fig.~\ref{fig:scaling}\,a for a fixed number $m$ and both energy
splittings shown in Fig.~\ref{fig:method_1}. While the equal-size
energy range splitting (Fig.~\ref{fig:method_1}\,a) is the most basic
approach, the run-time balanced energy splitting
(Fig.~\ref{fig:method_1}\,b) is chosen such that walkers in different
energy sub-windows complete the first WL iteration after the same
number of sweeps~(within statistical fluctuations). As the growth
behavior of WL histograms is in principle known~\cite{zhou05pre}, such
an energy splitting can be estimated by analyzing the first-iteration
histogram from a short pre-run with equal-size energy intervals.
Using the lipid system as an example and considering a much smaller
global energy range accessible for single-walker simulations, the
slope of speed-up in completing the first WL iteration is $\approx0.5$
for the equal size energy splitting and $\approx1.6$ for the run-time
balanced energy splitting, which is particularly remarkable as this
indicates that the speed-up is \emph{larger} than the number of
processors used.  For the HP protein (cf.  Fig.~\ref{fig:two_dos}\,b),
single walker WL simulations did not reach convergence of the DOS over
the entire energy range within a CPU year, yet all parallel runs
finalized within a month already for equal-size energy splitting and
with only a single walker per energy interval. We found that $s_{o =
  75\%}(h = 9) \approx 20$; again, we get a speed-up larger than the
number of processors even in this basic set-up.  To investigate the
weak scaling properties, we simulate the 10-state Potts model for
different system sizes. We increase the number of computing cores by
the same amount as the system size increases and measure the total run
time.  The results are shown in Fig.~\ref{fig:scaling}\,b), where
these data are compared to the run time increase for serial, single
walker WL simulations of the same model. Fig.~\ref{fig:scaling}
clearly shows that our method is able to achieve both, strong and weak
scaling, i.e., by increasing the number of computing cores one can get
results faster for the same system and/or simulate larger systems in
the same time.


To conclude, we introduced a generic, hierarchical parallel framework
for generalized ensemble WL simulations based on the concepts of
energy range splitting, replica exchange Monte Carlo and multiple
random walkers. The method is held as simple and general as possible
and leads to significant advantages over traditional, single-walker WL
sampling. In our complete formulation, we consider multiple WL walkers
in independent parallelization directions and show that strong
\textit{and} weak scaling can be achieved. (Our formulation far
surpasses a version with a single walker per equal-size energy
sub-window and an \textit{ad-hoc} overlap, which was used earlier to
study evaporation and condensation in a spin lattice
model~\cite{nogawa11pre}). With the ability to reach into previously
inaccessible domains, highly accurate results, and proven scalability
up to $\sim2000$ cores without introducing an erroneous bias, we
provide a proof of concept that our novel parallel WL scheme has the
potential for large scale parallel Monte Carlo simulations. Since the
framework is complementary to other technical parallelization
strategies, it is further extendible in a straightforward way. This
facilitates efficient simulations of larger and more complex problems
and thus provides a basis for many applications on petaflop
\hbox{machines}.

\begin{acknowledgments}
  This work is supported by the National Science Foundation under
  Grants DMR-0810223 and OCI-0904685. Y.W. Li was partly sponsored by
  the Office of Advanced Scientific Computing Research; U.S.
  Department of Energy. Part of the work was performed at the Oak
  Ridge Leadership Computing Facility at ORNL, which is managed by
  UT-Battelle, LLC under Contract No. De-AC05-00OR22725. Supercomputer
  time was provided by TACC under XSEDE grant PHY130009.
\end{acknowledgments}


\end{document}